\documentclass[preprint,prd,aps,showpacs,showkeys,nofootinbib]{revtex4}
\usepackage{graphicx}
\begin{document}


\title{Electroweak baryogenesis and electron EDM in the B-LSSM}

\author{Jin-Lei Yang$^{1,2,3}$\footnote{yangjinlei@itp.ac.cn},
Tai-Fu Feng$^{1,2,4}$\footnote{fengtf@hbu.edu.cn}, Hai-Bin Zhang$^{1,2}$\footnote{hbzhang@hbu.edu.cn}}

\affiliation{Department of Physics, Hebei University, Baoding, 071002, China$^1$\\
Hebei Key Lab of High-precision Computation and Application of Quantum Field Theory, Baoding, 071002, China$^2$\\
CAS Key Laboratory of Theoretical Physics, School of Physical Sciences, University of Chinese Academy of Sciences, Beijing 100049, China$^3$\\
Department of Physics, Chongqing University, Chongqing 401331, China$^4$}

\begin{abstract}
Electroweak baryogenesis (EWB) and electric dipole moment (EDM) have close relation with the new physics beyond the standard model (SM), because the SM CP-violating (CPV) interactions are not sufficient to provide the baryon asymmetry of the universe by many orders of magnitude, and the theoretical predictions for the EDM of electron ($d_e$) in the SM are too tiny to be detected in near future. In this work, we explore the CPV effects on EWB and the electron EDM in the minimal supersymmetric extension (MSSM) of the SM with local $B-L$ gauge symmetry (B-LSSM). And the two-step transition via tree-effects in this model is discussed. Including two-loop corrections to $d_e$ and considering the constrains from updated experimental data, the numerical results show that the B-LSSM can account for the observed baryon asymmetry. In addition, when the cancellation between different contributions to $d_e$ takes place, the region favored by EWB can be compatible with the corresponding EDM bound.

\end{abstract}


\maketitle

\section{Introduction\label{sec1}}
\indent\indent
Despite the considerable success of the Standard Model (SM) in describing a large amount of experimental observations, there are still various of evidences beyond the SM. One of the most interesting problems is the baryon asymmetry of the universe (BAU)~\cite{Cooke:2013cba,Ade:2015xua}:
\begin{eqnarray}
&&Y_B\equiv\frac{\rho_B}{s}=\left\{\begin{array}{l}
(8.2-9.4)\times10^{-11}\;\;\;(95\%{\rm CL}),\;\;{\rm Big\;Bang\;Nucleosynthesis}\\
8.65\pm0.09\times10^{-11},\qquad\quad\;\;\;\;\;\;\;\;{\rm PLANCK}\\
\end{array}\right.\;
\end{eqnarray}
where $\rho_B$ is the baryon number density, $s$ is the entropy density of the universe. The SM CP-violating (CPV) interactions are not sufficient to provide the asymmetry by many orders of magnitude, which indicates that the SM is incomplete. The search for new physics (NP) beyond the SM is motivated in part by the desire to overcome the failure of the SM to explain the BAU. Electroweak baryogenesis (EWB)~\cite{Kuzmin:1985mm} is an explanation of the origin of the cosmological asymmetry between matter and antimatter, and new CPV terms are needed to enhance the asymmetry theoretically.

Meanwhile, new CPV phases can provide much larger values of the electric dipole moments (EDMs) than the SM predictions. The SM prediction for the electron EDM is about $10^{-38}{\rm e\cdot cm}$~\cite{Gavela:1981sk,Bernreuther:1990jx,Pospelov:2013sca}, which is impossible to be detected by present experiments. However, when new CPV phases are introduced, the enhanced electron EDM may be detected in near future, which can be regarded as a smoking gun for NP beyond the SM. The upper bounds on $d_e$ have been obtained~\cite{de-1,PDG,de-3}
\begin{eqnarray}
&&|d_e|<8.7\times10^{-29}{\rm e\cdot cm}.
\end{eqnarray}
Since the experimental upper bound on the electron EDM is very small, the contributions from new CPV phases are limited strictly by the present experimental data, and researching NP effects on the electron EDM may shed light on the mechanism of CPV.

In extensions of the SM, the supersymmetry is considered as one of the most plausible candidates. And the analysis of EWB in the minimal supersymmetric extension of the SM (MSSM) are discussed in detail in Refs.~\cite{Dine:1990fj,Cohen:1992zx,Huet:1995sh,Chang:2002ex,Lee:2004we,Konstandin:2005cd,Chung:2008aya,Chung:2009qs,
Chung:2009cb,Cirigliano:2009yd,Chiang:2009fs,Morrissey:2012db,Kozaczuk:2012xv}, and in nonminimal supersymmetric models are discussed in Refs.~\cite{Pietroni:1992in,Davies:1996qn,Huber:2000mg,Kang:2004pp,Huber:2006wf}, which indicates that the main contributions to $Y_B$ come from the $T$-terms (the trilinear scalar terms in the soft supersymmetry breaking potential) and the $\mu$ term (the bilinear Higgs mass term in the superpotential). The supersymmetric effects on the EDM of electron has been explored in Refs.~\cite{Nath:1991dn,Kizukuri:1992nj,Falk:1996ni,Falk:1998pu,Brhlik:1998zn,Bartl:1999bc,Abel:2001vy,Barger:2001nu,
Olive:2005ru,Cirigliano:2006dg,YaserAyazi:2006zw,Engel:2013lsa,Chupp:2017rkp}. The results show that the most interesting possibility to suppress the electron EDM to below the corresponding experimental upper bound is, the contributions from different phases cancel each other. However, if we assume that the only CPV phases come from $\mu$ and $T$, the value of them is limited strictly by the experimental upper bounds on $d_e$. In a word, the CPV characters in supersymmetry are very interesting and studies on them may shed some light on the general characteristics of the supersymmetric model.

The MSSM with local $B-L$ gauge symmetry (B-LSSM)~\cite{Barger:2008wn,FileviezPerez:2008sx,5,6} is based on the gauge symmetry group $SU(3)\otimes SU(2)_L\otimes U(1)_Y\otimes U(1)_{B-L}$, where $B$ stands for the baryon number and $L$ stand for the lepton number respectively. Compared with the MSSM,  B-LSSM can provide much more candidates for the dark matter~\cite{16,1616,DelleRose:2017ukx,DelleRose:2017uas}, and the invariance under $U(1)_{B-L}$ gauge group imposes the R-parity conservation which is assumed in the MSSM to avoid proton decay. In addition, the model also alleviates the little hierarchy problem of the MSSM~\cite{search,77,88,9,99,10,11}. In this paper, we explore the CPV effects on $Y_B$ and the electron EDM $d_e$ in the B-LSSM. And the possible cancellation between different contributions to $d_e$ is explored, which is different from the case in the MSSM. Compared with the MSSM, there are new CPV terms in the B-LSSM, the cancellation between the contributions to $d_e$ from these new CPV phases and the phase of $M_1$ which is the phase of gaugino mass term, in this paper. Moreover, there are two mass terms which can be small and make contributions to $d_e$, the effects of them are also explored in detail.

The paper is organized as follows. In Sec.II, we describe the B-LSSM briefly by introducing the superpotential and the general soft breaking terms. Then the analysis on electroweak phase transition (PT), $Y_B$ and the electron EDM $d_e$ in the B-LSSM are presented in Sec.III. In Sec.IV, we explore the CPV effects on $Y_B$, $d_e$ by varying different parameters. Conclusions are summarized in Sec.V.

\section{The B-LSSM\label{sec2}}

In the B-LSSM, two chiral singlet superfields $\hat{\eta}_{1}\sim(1,1,0,-1)$, $\hat{\eta}_{2}\sim(1,1,0,1)$ and three generations of right-handed neutrinos are introduced, which allow for a spontaneously broken $U(1)_{B-L}$ without necessarily breaking R-parity. In addition, this version of B-LSSM is encoded in SARAH~\cite{164}, which is used to create the mass matrices and interaction vertexes in the model. Meanwhile, the superpotential of the B-LSSM can be written as
\begin{eqnarray}
&&W=Y_u^{ij}\hat{Q_i}\hat{H_2}\hat{U_j^c}+\mu \hat{H_1} \hat{H_2}-Y_d^{ij} \hat{Q_i} \hat{H_1} \hat{D_j^c}
-Y_e^{ij} \hat{L_i} \hat{H_1} \hat{E_j^c}+\nonumber\\
&&\;\;\;\;\;\;\;\;\;Y_{\nu, ij}\hat{L_i}\hat{H_2}\hat{\nu}^c_j-\mu' \hat{\eta}_1 \hat{\eta}_2
+Y_{x, ij} \hat{\nu}_i^c \hat{\eta}_1 \hat{\nu}_j^c,
\end{eqnarray}
where $i, j$ are generation indices. Then the soft breaking terms of the B-LSSM are generally given as
\begin{eqnarray}
&&\mathcal{L}_{soft}=\Big[-\frac{1}{2}(M_1\tilde{\lambda}_{B} \tilde{\lambda}_{B}+M_2\tilde{\lambda}_{W} \tilde{\lambda}_{W}+M_3\tilde{\lambda}_{g} \tilde{\lambda}_{g}+2M_{BB'}\tilde{\lambda}_{B'} \tilde{\lambda}_{B}+M_{B'}\tilde{\lambda}_{B'} \tilde{\lambda}_{B'})-
\nonumber\\
&&\hspace{1.4cm}
B_\mu H_1H_2 -B_{\mu'}\tilde{\eta}_1 \tilde{\eta}_2 +T_{u,ij}\tilde{Q}_i\tilde{u}_j^cH_2+T_{d,ij}\tilde{Q}_i\tilde{d}_j^cH_1+
T_{e,ij}\tilde{L}_i\tilde{e}_j^cH_1+T_{\nu}^{ij} H_2 \tilde{\nu}_i^c \tilde{L}_j+\nonumber\\
&&\hspace{1.4cm}
T_x^{ij} \tilde{\eta}_1 \tilde{\nu}_i^c \tilde{\nu}_j^c+h.c.\Big]-m_{\tilde{\nu},ij}^2(\tilde{\nu}_i^c)^* \tilde{\nu}_j^c-
m_{\tilde{q},ij}^2\tilde{Q}_i^*\tilde{Q}_j-m_{\tilde{u},ij}^2(\tilde{u}_i^c)^*\tilde{u}_j^c-m_{\tilde{\eta}_1}^2 |\tilde{\eta}_1|^2-\nonumber\\
&&\hspace{1.4cm}
m_{\tilde{\eta}_2}^2 |\tilde{\eta}_2|^2-m_{\tilde{d},ij}^2(\tilde{d}_i^c)^*\tilde{d}_j^c-m_{\tilde{L},ij}^2\tilde{L}_i^*\tilde{L}_j-
m_{\tilde{e},ij}^2(\tilde{e}_i^c)^*\tilde{e}_j^c-m_{H_1}^2|H_1|^2-m_{H_2}^2|H_2|^2,
\end{eqnarray}
where $\tilde\lambda_{B}, \tilde\lambda_{B'}$ denoting the gaugino of $U(1)_Y$ and $U(1)_{(B-L)}$ respectively. The local gauge symmetry $SU(2)_L\otimes U(1)_Y\otimes U(1)_{B-L}$ breaks down to the electromagnetic symmetry $U(1)_{em}$ as the Higgs fields receive vacuum expectation values (VEVs):
\begin{eqnarray}
&&H_1^1=\frac{1}{\sqrt2}(v_1+{\rm Re}H_1^1+i{\rm Im}H_1^1),
\qquad\; H_2^2=\frac{1}{\sqrt2}(v_2+{\rm Re}H_2^2+i{\rm Im}H_2^2),\nonumber\\
&&\tilde{\eta}_1=\frac{1}{\sqrt2}(u_1+{\rm Re}\tilde{\eta}_1+i{\rm Im}\tilde{\eta}_1),
\qquad\;\quad\;\tilde{\eta}_2=\frac{1}{\sqrt2}(u_2+i{\rm Re}\tilde{\eta}_2+i{\rm Im}\tilde{\eta}_2)\;.
\end{eqnarray}
For convenience, we define $u^2=u_1^2+u_2^2,\; v^2=v_1^2+v_2^2$ and $\tan\beta^{'}=\frac{u_2}{u_1}$ in analogy to the ratio of the MSSM VEVs ($\tan\beta=\frac{v_2}{v_1}$).

New $U(1)_{B-L}$ gauge group introduces new gauge boson $Z'$ and the corresponding gauge coupling constant $g_{_B}$. In addition, two Abelian groups gives rise to a new effect absent in the MSSM or other SUSY models with just one Abelian gauge group: the gauge kinetic mixing. Immediate interesting consequence of the gauge kinetic mixing arise in various sectors of the model. Firstly, new gauge boson $Z'$ mixes with the $Z$ boson in the MSSM, and new gauge coupling constant $g_{_{YB}}$ is introduced. Then the gauge kinetic mixing leads to the mixing between the $H_1^1,\;H_2^2,\;\tilde{\eta}_1,\;\tilde{\eta}_2$ at the tree level, and $\tilde\lambda_{B'}$ mixes with the two higgsinos in the MSSM at the tree level. Meanwhile, additional D-terms contribute to the mass matrices of the squarks and sleptons. All of these properties affect the theoretical predictions for $Y_B$ and $d_e$ in the B-LSSM, and the model are introduced in detail in our earlier work~\cite{Yang:2018utw,JLYang:2018,Yang:2018guw}.

\section{EWB and electron EDM in the B-LSSM\label{sec3}}

\subsection{Electroweak phase transion\label{secA}}

In the MSSM, EWB has been excluded because the strong first order PT with very light right handed stop $<120{\rm GeV}$ is not possible after the discovery of the $125\;{\rm GeV}$ Higgs boson~\cite{Delepine:1996vn,Carena:1996wj,Espinosa:1996qw,Carena:1997ki,Huber:1998ck,Carena:2002ss,Profumo:2007wc,Carena:2008vj,
Curtin:2012aa,Krizka:2012ah}. With respect to the MSSM, a strong two-step PT can be achieved in the B-LSSM, because there are two additional scalar singlets. These new singlets mix with the two doublets in the MSSM at the tree level through gauge kinetic mixing, which change the effective potential vastly. For simplicity, the temperature dependence of $\beta$, $\beta'$ is neglected and the tree-level effective potential can be written as
\begin{eqnarray}
&&V_{eff}(h,\eta)=\frac{1}{2}M(T)^2 h^2+\frac{1}{2}m_{\eta}^2 \eta^2+\frac{1}{32}(g1^2+g2^2+g_{_{YB}}^2)c_{2\beta}^2h^4+\frac{1}{8}g_{_B}^2c_{2\beta'}^2\eta^4\nonumber\\
&&\qquad\qquad\quad+\frac{1}{8}g_{_B}g_{_{YB}}c_{2\beta}c_{2\beta'}h^2\eta^2,\label{Veff}
\end{eqnarray}
where
\begin{eqnarray}
&&M(T)^2\equiv M_0^2+\mathcal{G}T^2=m_{H_1}^2c_{\beta}^2+m_{H_2}^2s_{\beta}^2+2\mu^2-B_\mu s_{2\beta}+\mathcal{G}T^2,\label{MT}\\
&&m_\eta^2=m_{\eta_1}^2 c_{\beta'}^2+m_{\eta_2}^2 s_{\beta'}^2+2\mu_\eta^2-B_\eta s_{2\beta'},\\
&&c_\beta\equiv\cos\beta,\;s_\beta\equiv\sin\beta,\;c_{2\beta}\equiv\cos2\beta,\;s_{2\beta}\equiv\sin2\beta,
\end{eqnarray}
and $T$ denotes temperature, $\mathcal{G}$ is the sum of relevant couplings, $h$ and $\eta$ acquire VEVs $<h>=v$, $<\eta>=u$ respectively at zero temperature (present universe). Since the singlets couples to fewer degrees of freedom, their thermal masses is lower than that of the SM higgs, and we ignore their thermal mass. At very high temperature, $h$ and $\eta$ are stabilized at the origin. In addition, it can be noted in Eq. (\ref{Veff},$\;$\ref{MT}) that, the only possible gauge-dependence term is $\mathcal{G}T^2$, and the gauge-independence of $\mathcal{O}(T^2)$ term was proved in the appendix C of Ref.~\cite{Patel:2011th}. Hence our analysis of the electroweak PT is gauge invariant. As the universe cools, the singlets transition to a nonzero VEV $u_{c1}$ first, in a second order phase transition at $T_{c1}$. Then at temperature $T_{c2}\sim m_W<T_{c1}$, the universe undergoes a first order PT to $(v_{c2},u)$. Then we can obtain $v_{c2}$ and $M(T)^2$ by solving the equations
\begin{eqnarray}
&&\left\{\begin{array}{l}
V_{eff}(0,u_{c1}){\Big|}^{T_{c1}}=V_{eff}(v_{c2},u){\Big|}^{T_{c2}},\\
\\
\frac{\partial V_{eff}}{\partial h}{\Big|}^{T_{c2}}_{(v_{c2},u)}=0.
\end{array}\right.\;
\end{eqnarray}
Then the first order transition temperature $T_{c2}$ can be obtained by $\sqrt{(M(T)^2-M_0^2)/\mathcal{G}}$. For the EWB to work, the sphaleron process must be decoupled when the electroweak PT completes. In other words, the sphaleron rate in the broken phase should be less than the Hubble parameter at that moment. In general, the sphaleron decoupling condition is cast into the form $v_{c2}/T_{c2}\gtrsim1$. And in our chosen parameter space in the next section, we have $v_{c2}/T_{c2}\gtrsim1.5$, which is sufficient for the taking place of EWB.

\subsection{Baryon asymmetry $Y_B$\label{secB}}
The CPV effects enter as source terms in the quantum transport equations that govern the production of chiral charge at the phase boundary. According to Ref.~\cite{Lee:2004we}, a simple expression for the baryon-to-entropy ratio can be written as
\begin{eqnarray}
&&Y_B=-F_1\sin \theta_\mu+F_2\sin (-\theta_\mu+\theta_T).
\end{eqnarray}
where we have taken the gaugino mass terms $M_{1,2,B'}$ to be real, $\theta_\mu$, $\theta_T$ are the phases of $\mu$ and $T_e$ respectively. Compared with the expression in Ref.~\cite{Lee:2004we}, the additional minus sign on $\theta_\mu$ comes from different definition of $\mu$. The coefficients $F_i$ depend on the mass parameters in the B-LSSM, such as $\mu$, $M_1$, $M_2$, $M_{B'}$, $A_0$ (we assume that T-terms are all same at the GUT scale, $T_u/Y_u=T_d/Y_d=T_e/Y_e=A_0$, where $Y_{u,d,e}$ are the corresponding Yukawa coupling constants) and squark masses $M_{\tilde t_L}$, $M_{\tilde t_R}$. In addition, $F_i$ also have a overall dependence on bubble wall parameters $v_w$, $L_w$, $\Delta\beta$. For the concrete expressions of $F_i$, we adopt the formulas displayed in Ref.~\cite{Lee:2004we}. Compared with the MSSM, there is new contribution to $S_{\tilde H}^{{{CP}\!\!\!\!\!\!/}}$ (the CPV higgsino source) in the B-LSSM, which comes from the mixing between new gaugino $\tilde\lambda_{B'}$ and the two higgsinos in the MSSM through gauge kinetic mixing, and the corresponding gauge coupling constant is $g_{_{YB}}$.

\subsection{The EDM of electron $d_e$\label{secC}}
\begin{figure}
\setlength{\unitlength}{1mm}
\centering
\includegraphics[width=5in]{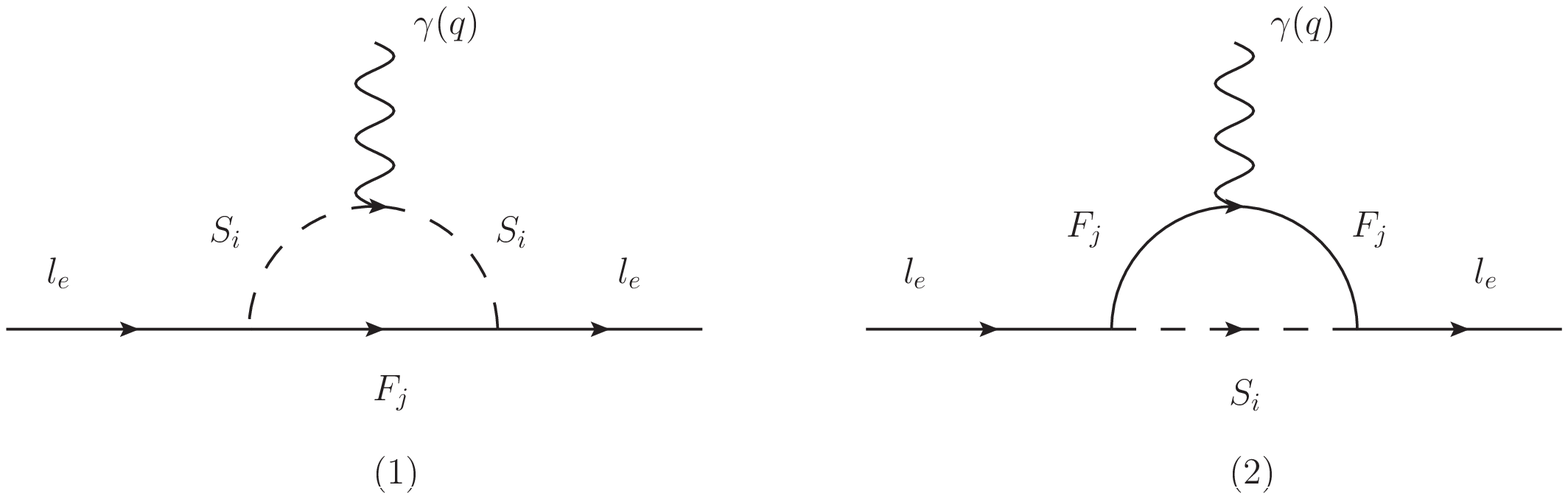}
\vspace{0cm}
\caption[]{The one-loop level diagrams contribute to the electron EDM, where (a) represents the charged scalars loops, and (b) represents the charged fermions loops.}
\label{Feynman diagram}
\end{figure}

\begin{figure}
\setlength{\unitlength}{1mm}
\centering
\includegraphics[width=6in]{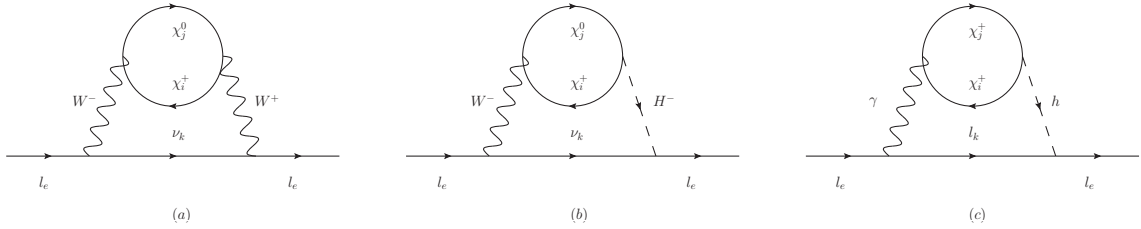}
\vspace{0cm}
\caption[]{The two-loop Barr-Zee type diagrams in which a closed fermion loop is attached to the virtual gauge bosons or Higgs fields, the corresponding contributions to $d_e$ are obtained by attaching a photon to the internal particles in all possible ways.}
\label{Feynman diagram two-loop}
\end{figure}

The effective Lagrangian for the electron EDM can be written as
\begin{eqnarray}
&&\mathcal{L}_{EDM}=-\frac{i}{2}d_e\bar l_e\sigma^{\mu\nu}\gamma_5 l_e F_{\mu\nu}.
\end{eqnarray}
where $\sigma^{\mu\nu}=i[\gamma^\mu,\gamma^\nu]/2$, and $F_{\mu\nu}$ is the electromagnetic field strength. Adopting the effective Lagrangian approach, we can get
\begin{eqnarray}
&&d_e=-\frac{2eQ_fm_e}{(4\pi)^2}\Im(C_2^R+C_2^{L*}+C_6^R),
\label{EDMe}
\end{eqnarray}
where $Q_f=-1$, $m_e$ denotes the electron mass, and $C_{2,6}^{L,R}$ represent the Wilson coefficients of the corresponding operators $O_{2,6}^{L,R}$
\begin{eqnarray}
&&O_2^{L,R}=\frac{eQ_f}{(4\pi)^2}(-iD_\alpha^*) \bar l_e \gamma^\alpha F\cdot\sigma P_{L,R}l_e,\nonumber\\
&&O_6^{L,R}=\frac{eQ_fm_e}{(4\pi)^2}\bar l_e F\cdot\sigma P_{L,R}l_e,
\end{eqnarray}
where $D_\alpha=\partial_\alpha+i A_\alpha$, $l_e$ is the wave function for electron, and $P_{R,L}=(1\pm\gamma_5)/2$. Then, the Feynman diagrams contributing to the above Wilson coefficients are depicted by Fig.~\ref{Feynman diagram}.
Calculating the Feynman diagrams, the electron EDM can be written as
\begin{eqnarray}
&&d_e^{(1)}=\frac{-2}{em_e}\Im\Big\{x_e[-I_3(x_{F_j},x_{S_i})+I_4(x_{F_j},x_{S_i})][(C_{\bar l_e S_i F_j}^LC_{\bar F_j S_i l_e}^R)+(C_{\bar l_e S_i F_j}^RC_{\bar F_j S_i l_e}^L)^*]\nonumber\\
&&\qquad\quad+\sqrt{x_{e}x_{F_j}}[-2I_1(x_{F_j},x_{S_i})+2I_3(x_{F_j},x_{S_i})]C_{\bar l_e S_i F_j}^RC_{\bar F_j S_i l_e}^R\Big\},\nonumber\\
&&d_e^{(2)}=\frac{-2}{em_e}\Im\Big\{x_e[-I_1(x_{F_j},x_{S_i})+2I_3(x_{F_j},x_{S_i})-I_4(x_{F_j},x_{S_i})][(C_{\bar l_e S_i F_j}^RC_{\bar F_j S_i l_e}^L)\nonumber\\
&&\qquad\quad+(C_{\bar l_e S_i F_j}^LC_{\bar F_j S_i l_e}^R)^*]+\sqrt{x_{e}x_{F_j}}[2I_1(x_{F_j},x_{S_i})-2I_2(x_{F_j},x_{S_i})-2I_3(x_{F_j},x_{S_i})]\nonumber\\
&&\qquad\quad\times C_{\bar l_e S_i F_j}^RC_{\bar F_j S_i l_e}^R\Big\},
\end{eqnarray}
where $x_i=m_i^2/m_W^2$, $C_{abc}^{L,R}$ denotes the constant parts of the interactional vertex about $abc$, which can be got through SARAH, $a$, $b$, $c$ denote the interactional particles, and the concrete expressions for the functions $I_{1,2,3,4}$ can be found in~\cite{Zhang:2013hva,Zhang:2013jva}. In addition, our earlier work~\cite{Yang:2018guw} shows that, two-loop Barr-Zee type diagrams can make important contributions to the muon magnetic dipole moment (MDM), and  we consider the contributions from the two-loop diagrams in which a closed fermion loop is attached to the virtual gauge bosons or Higgs fields. According to Ref.~\cite{Yang:2009zzh}, the main two-loop diagrams contributing to the electron EDM are shown in Fig.~\ref{Feynman diagram two-loop}.
Assuming $m_F=m_{\chi^+_i}=m_{\chi^0_j}\gg m_W$, $m_F=m_{\chi^+_i}\gg m_{h}$, the contributions from the two-loop diagrams to $d_e$ can be simplify as
\begin{eqnarray}
&&d_e^{(a)}=\frac{3G_F m_W\sqrt{x_{e}}}{-64\sqrt{2}\pi^4}\Big\{\Im(C_{\bar f_jWf_i}^LC_{\bar f_jWf_i}^{R*})\Big\},\nonumber\\
&&d_e^{(b)}=\frac{G_F e m_{W} C_{\bar l_e H \nu_k}^L}{256\pi^4 g_2\sqrt{x_{_F}}}\Big\{\Big[179/36+10/3J(x_{_F},x_{_W},x_{_H})\Big]\Im(C_{\bar f_iHf_j}^L
C_{\bar f_jWf_i}^L+C_{\bar f_iHf_j}^RC_{\bar f_jW\chi_i^+}^R)\nonumber\\
&&\qquad\quad+\Big[-1/9-2/3
J(x_{_F},x_{_W},x_{_H})\Big]\Im(C_{\bar f_iHf_j}^LC_{\bar f_jWf_i}^R+C_{\bar f_iHf_j}^RC_{\bar f_jWf_i}^L)\nonumber\\
&&\qquad\quad+\Big[-16/9-8/3
J(x_{_F},x_{_W},x_{_H})\Big]\Im(C_{\bar f_iHf_j}^LC_{\bar f_jWf_i}^L-C_{\bar f_iHf_j}^RC_{\bar f_jW\chi_i^+}^R)\nonumber\\
&&\qquad\quad+\Big[-2/9-4/3
J(x_{_F},x_{_W},x_{_H})\Big]\Im(C_{\bar f_iHf_j}^LC_{\bar f_jWf_i}^R-C_{\bar f_iHf_j}^RC_{\bar f_jWf_i}^L)\Big\},\nonumber\\
&&d_e^{(c)}=\frac{G_F e m_{W} C_{\bar l_e h^0 l_e}}{64\pi^4 \sqrt{x_{_F}}}\Im(C_{\bar f_i h^0 f_i}^L)\Big[1+\ln\frac{x_{_F}}{x_{_h}}\Big],
\end{eqnarray}
where $f_j$, $f_i$ denote $\chi_j^0$ and $\chi_i^\pm$ respectively, $W$ denotes $W$ boson, $H$ denotes charged Higgs boson, $h$ denotes SM-like Higgs boson, the concrete expressions for the function $J$ can be found in Ref.~\cite{Yang:2018guw}.

Compared with the MSSM, there are new CPV terms $M_{BB'}$, $M_{B'}$ and $\mu'$ can contribute to the electron EDM. In the next section, we will explore the possible cancellation between the contributions to $d_e$ from these new CPV terms and $M_1$. Moreover, it is more interesting that there are two new mass terms in the B-LSSM, the mixing mass term $M_{BB'}$ between $\tilde\lambda_{B}$, $\tilde\lambda_{B'}$, and $M_{B'}$ which is the mass term of $\tilde\lambda_{B'}$. Both of $M_{BB'}$ and $M_{B'}$ can be very small and the gaugino masses still can be large enough to satisfy the experimental upper bounds on gaugino masses. The contributions to the electron EDM from the phases of $M_{BB'}$ or $M_{B'}$ can be highly suppressed by small $M_{BB'}$ or $M_{B'}$ to satisfy the present experimental upper bound on $d_e$.


\section{Numerical analyses\label{sec4}}

In this section, we present the numerical results of $Y_B$ and electron EDM $d_e$ in the B-LSSM. The EDMs of neutron, mercury, heavy quarks are discussed in our previous work~\cite{Yang:2019aao}, and some two-loop Barr-Zee and gluino type corrections are considered, the numerical results show that the constraints from these quantities are less strict than the constraints from the electron EDM. Hence, the allowed regions by the present upper limit on $d_e$ can coincide with the upper limits on EDMs of neutron, mercury and heavy quarks. The relevant SM input parameters are chosen as $m_W=80.385\;{\rm GeV},\;m_Z=90.1876\;{\rm GeV},\;\alpha_{em}(m_Z)=1/128.9,\;\alpha_s(m_Z)=0.118$. We take $Y_\nu=Y_x=0$ approximately due to the tiny neutrino masses basically do not affect the numerical analysis. The SM-like Higgs boson mass is $125.18\;{\rm GeV}$~\cite{PDG} and constrains the parameter space strictly. Compared with the MSSM, new singlets mix with the MSSM doublets at the tree level in the B-LSSM, which can affect the theoretical prediction of SM-like Higgs mass. Including the leading-log radiative corrections from stop and top particles~\cite{HiggsC1,HiggsC2,HiggsC3}, the lightest Higgs mass in the B-LSSM is limited in the range $124\;{\rm GeV}<m_h<126\;{\rm GeV}$ in our chosen parameter space below.

The updated experimental data~\cite{newZ} on searching $Z'$ indicates $M_{Z'}\geq4.05\;{\rm TeV}$ at $95\%$ Confidence Level (CL). And an upper bound on the ratio between the $Z'$ mass and its gauge coupling is given in Refs.~\cite{20,21} at $99\%$ CL as $M_{Z'}/g_B>6\;{\rm TeV}$. In our earlier work~\cite{Yang:2018guw}, we explore the effects of parameters $\tan\beta$, $\tan\beta'$, $g_{_B}$, $g_{_{YB}}$ and slepton masses $M_{\tilde L,\tilde e}$ on the muon MDM in the B-LSSM without CPV. Since the CPV phases affect the electron EDM more obviously than the muon MDM, we explore the CPV effects on the electron EDM firstly in this paper, but put off the exploration of CPV effects on the muon MDM in our next work. Then considering the experimental data of the muon MDM, we choose $M_{Z'}=4.2\;{\rm TeV}$, $\tan\beta=10$, $\tan\beta'=1.15$, $g_B=0.4$, $M_{\tilde L,\tilde e}={\rm diag}(2,2,2)\;{\rm TeV}$. We don't fix $g_{_{YB}}$ because it affects the numerical result of $Y_B$ obviously through the contributions from new gaugino $\tilde\lambda_{B'}$. Considering the constraints from the experiments~\cite{PDG}, for those parameters in higgsino and gaugino sectors, we appropriately fix $M_1=\frac{1}{2}M_2=\frac{1}{2}M_{B'}=\frac{1}{2}M_{BB'}=0.3\;{\rm TeV}$, $\mu'=0.8\;{\rm TeV}$ for simplicity. The value of $\mu$ is not fixed, because the main contributions to $Y_B$ come from the $\mu$ term. In addition, in order to satisfy the experimental data on $\bar B\rightarrow X_s\gamma$ and $B_s^0\rightarrow \mu^+\mu^-$~\cite{JLYang:2018}, we take the stop mass $m_{\tilde t_L}=m_{\tilde t_R}=1.5\;{\rm TeV}$, charged Higgs boson mass $M_{H^\pm}=1.5\;{\rm TeV}$ for simplicity. According to Refs.~\cite{AlvarezGaume:1983gj,Strumia:1996pr,Profumo:2006yu}, the size of the scalar trilinear
couplings are limited by the conditions of avoiding charge and color breaking minima, then we can take $A_{0}=0.1\;{\rm TeV}$, which can satisfy this condition. For the bubble wall parameters $v_w$, $L_w$, we adopt the central values $v_w=0.05$, $L_w=25/T$~\cite{Moreno:1998bq,John:2000zq}, and $\Delta\beta$ as a function of pseudoscalar Higgs boson mass provided in Ref.~\cite{Moreno:1998bq}, we take $\Delta\beta=0.015$. For the thermal widths, we adopt the results in Ref.~\cite{Akula:2017yfr} in the following analysis.

\begin{figure}
\setlength{\unitlength}{1mm}
\centering
\includegraphics[width=3.1in]{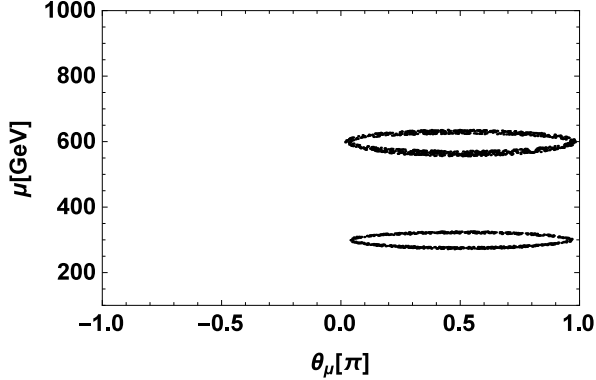}
\vspace{0cm}
\caption[]{Keeping $Y_B$ in the region $(8.2-9.4)\times10^{-11}$, the allowed region of $\theta_\mu$ and $\mu$.}
\label{YBtheta}
\end{figure}

In order to see how $\theta_\mu$, $\theta_{A_0}$ and $\mu$ affect $Y_B$, we take $g_{_{YB}}=-0.4$ and scan the regions of the parameter space [$\theta_\mu=(-\pi,\pi),\;\theta_{A_0}=(-\pi,\pi),\;\mu=(0.1,1)\;{\rm TeV}$]. In the scanning, we keep $Y_B$ in the region $(8.2-9.4)\times10^{-11}$. Then the allowed region of $\theta_\mu$ and $\mu$ is displayed in Fig.~\ref{YBtheta}. From the picture, we can see that there are two ellipses in the figure, and the two ellipses mainly concentrate on the vicinity of $\mu=600\;{\rm GeV}$ and $\mu=300\;{\rm GeV}$ respectively, because the effects of these interactions are resonantly enhanced when $\mu$ is comparable to the mass terms $M_{1,2,B'}$~\cite{Carena:1997gx,Carena:2000id}, the observed baryon asymmetry can be accounted for only in this case. The allowed region of $\theta_\mu$ is concentrated on $\theta_\mu>0$, because the mainly contributions come from the coefficient $F_1$, and $F_1$ is negative in our chosen parameter space. In addition, with the increasing of $\theta_\mu$, the value of $\mu$ has a small deviation from $M_{1,B'}$ or $M_2$, because the contributions to $Y_B$ with large $\theta_\mu$ will exceed $9.4\times10^{-11}$ when $\mu$ equals to $M_{1,B'}$ or $M_2$. It also can be noted from picture that, the minimum value of $\theta_\mu$ is about $0.03$ for $\mu=600\;{\rm GeV}$, $0.04$ for $\mu=300\;{\rm GeV}$, and there are more points in the vicinity of $\mu=600\;{\rm GeV}$. In our chosen parameter space, the value of $M_{B'}$ is $600\;{\rm GeV}$, and there is also a resonant enhancement when $\mu=M_{B'}$. Hence, there are more points in the vicinity of $\mu=600\;{\rm GeV}$ and the minimum value of $\theta_\mu$ is smaller slightly for $\mu=600\;{\rm GeV}$ than $\mu=300\;{\rm GeV}$.

\begin{figure}
\setlength{\unitlength}{1mm}
\centering
\includegraphics[width=3.1in]{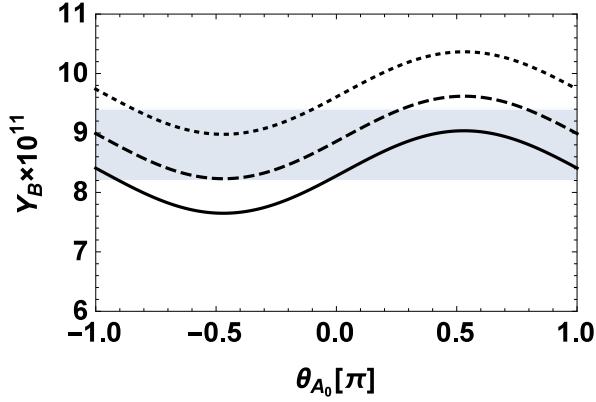}
\vspace{0cm}
\caption[]{$Y_B$ versus $\theta_{A_0}$ for $g_{_{YB}}=-0.3$ (solid line), $-0.4$ (dashed line), $-0.5$ (dotted line), and the gray area denotes the experimental interval $(8.2-9.4)\times10^{-11}$.}
\label{YB-T}
\end{figure}
Then we take $\theta_\mu=0.03$, $\mu=600\;{\rm GeV}$, and explore how $\theta_{A_0}$ and new parameter $g_{_{YB}}$ affect the numerical result. $Y_B$ versus $\theta_{A_0}$ for $g_{_{YB}}=-0.3$ (solid line), $-0.4$ (dashed line), $-0.5$ (dotted line) is plotted in Fig.~\ref{YB-T}, where the gray area denotes the experimental interval $(8.2-9.4)\times10^{-11}$. Compared with the MSSM, new parameter $g_{_{YB}}$ can affect the numerical result obviously, and $Y_B$ increases with the increasing of $|g_{_{YB}}|$, because the contribution from new gaugino $\tilde\lambda_{B'}$ is proportional to $g_{_{YB}}^2$~\cite{Lee:2004we}. From the picture we can see that, in our chosen parameter space, the allowed region of $\theta_{A_0}$ is larger when $g_{_{YB}}=-0.4$ than $g_{_{YB}}=-0.3$ or $-0.5$. The value of $g_{_{YB}}$ under which the allowed region of $\theta_{A_0}$ is largest depends on the value of $\theta_\mu$ and $\mu$, because the main contributions to $Y_B$ come from the $\mu$ term.

\begin{figure}
\setlength{\unitlength}{1mm}
\centering
\includegraphics[width=3.1in]{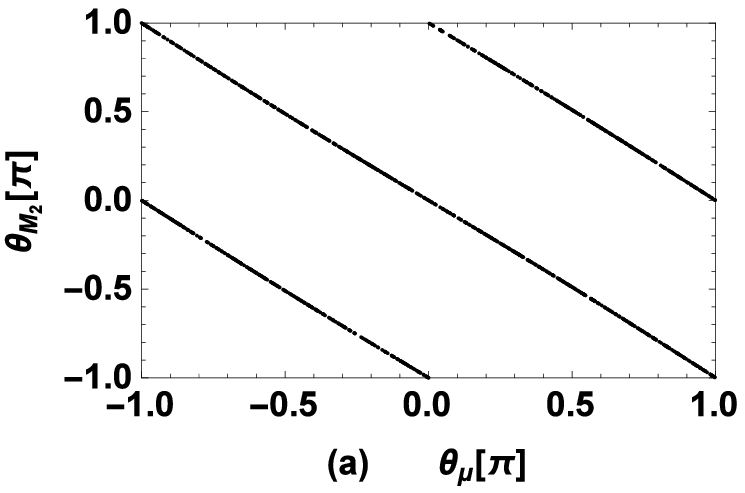}%
\vspace{0.5cm}
\includegraphics[width=3.1in]{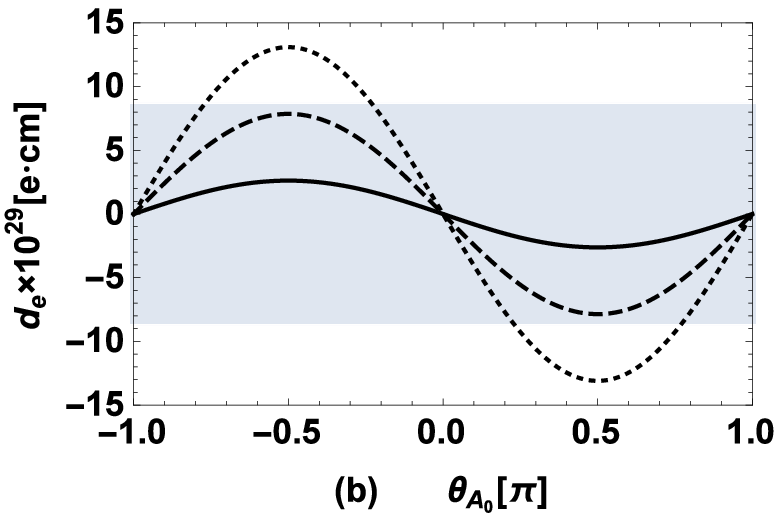}
\vspace{0cm}
\caption[]{Keeping $|d_e|<8.7\times10^{-29}$, the cancellation between $\theta_{M_2}$ and $\theta_\mu$ (a) are shown. And $d_e$ versus $\theta_{A_0}$ are plotted for $A_0=0.1\;{\rm TeV}$ (solid line), $0.3\;{\rm TeV}$ (dashed line), $0.5\;{\rm TeV}$ (dotted line), where the gray area denotes the present experimental upper bound on $d_e$}
\label{de-theta12}
\end{figure}
From the numerical results of $Y_B$, we can see that the minimum value of $\theta_\mu$ is about $0.03$ when EWB can take place. However, in this case, the contributions to the electron EDM $d_e$ are enhanced vastly, and $d_e$ exceeds the corresponding upper bound by several orders of magnitude. Hence, the contributions from different CPV phases should cancel each other to satisfy the present experimental upper bound. Then we take $g_{_{YB}}=-0.4$, and explore the cancellation between $\theta_{M_2}$ and $\theta_\mu$ by taking other CPV phases equal to $0$. We scan the regions of the parameter space [$\theta_{M_2}=(-\pi,\pi),\;\theta_\mu=(-\pi,\pi)$], and keep $|d_e|<8.7\times10^{-29}$ in the scanning. The numerical results are shown in Fig.~\ref{de-theta12} (a). In addition, $\theta_{A_0}$ can also make important contributions to $Y_B$, hence it is interesting to explore how $\theta_{A_0}$ affects $d_e$. Since the effects of $\theta_{A_0}$ are highly suppressed by small $Y_e$, we do not have to cancel the contributions from $\theta_{A_0}$ to $d_e$. Then we plot $d_e$ versus $\theta_{A_0}$ in Fig.~\ref{de-theta12} (b), where the gray area denotes the experimental upper bound on $d_e$, the solid, dashed and dotted lines denote $A_0=0.1\;{\rm TeV},\;0.3\;{\rm TeV},\;0.5\;{\rm TeV}$, respectively.

From the pictures we can see that, the contributions from $\theta_{M_2}$ and $\theta_\mu$ are cancelled when $\theta_{M_2}\approx-\theta_\mu+n\pi$ ($n=0,\pm1$), the phases we chosen to cancel each other due to that, the contributions from $\theta_{M_2}$ and $\theta_\mu$ are comparable. In addition, the contributions from $\theta_{A_0}$ are enlarged by large $A_0$, and the contributions from $\theta_{M_2}$, $\theta_\mu$ are lager than $\theta_{A_0}$ by several orders of magnitude, hence the contributions from $\theta_\mu$ are hardly cancelled by $\theta_{A_0}$ (when the cancellation between $\theta_\mu$ and $\theta_{A_0}$ takes place, the maximum value of $\theta_\mu$ is ${\mathcal O}(10^{-3})$, which is not sufficient for the taking place of EWB). It is different from the case in the MSSM~\cite{YaserAyazi:2006zw}, in which the maximum value of $\theta_\mu$ can be large enough to the taking place of EWB, when the cancellation happens between the contributions from $\theta_\mu$ and $\theta_{A_0}$ to $d_e$. It results from that, the contributions from sleptons are highly suppressed by large slepton masses, in our chosen parameter space. It can be noted that, the contributions from $\theta_\mu$ to $d_e$ can be cancelled by $\theta_{M_2}$, and $\theta_\mu$ is the main source of baryon asymmetry, hence the worry about the contributions from the large value of $\theta_\mu$, which is needed to give rise to EWB, to the electron EDM whether can be cancelled is relaxed.

\begin{figure}
\setlength{\unitlength}{1mm}
\centering
\includegraphics[width=3.1in]{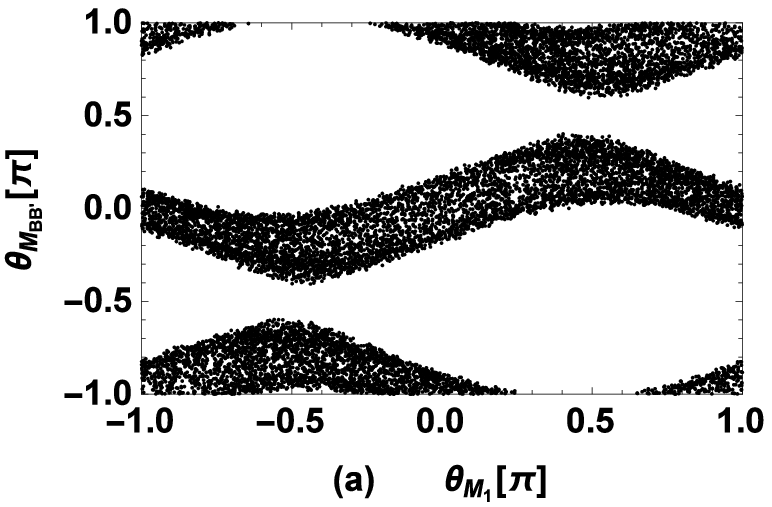}%
\vspace{0.5cm}
\includegraphics[width=3.1in]{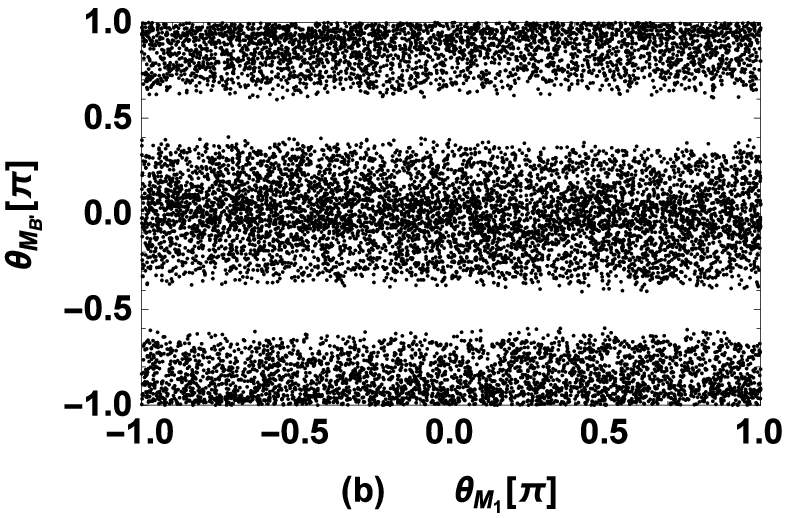}
\vspace{0cm}
\caption[]{Keeping $|d_e|<8.7\times10^{-29}$, when the possible cancellation between $\theta_{M_1}$, $\theta_{\mu'}$, $\theta_{M_{BB'}}$, $\theta_{M_{B'}}$ take place, the allowed regions of $\theta_{M_{BB'}}$, $\theta_{M_1}$ (a) and $\theta_{M_{B'}}$, $\theta_{M_1}$ (b) are plotted.}
\label{de-thetanew}
\end{figure}

In the B-LSSM, there are new CPV phases $\theta_{\mu'}$, $\theta_{M_{BB'}}$ and $\theta_{M_{B'}}$ can make contributions to the electron EDM. In addition, the gaugino mass term $M_1$ can also have CPV phase $\theta_{M_1}$, and makes contributions to $d_e$. Then we set other CPV phases equal to zero and explore the cancellation between $\theta_{\mu'}$, $\theta_{M_{BB'}}$, $\theta_{M_{B'}}$ and $\theta_{M_1}$. Scanning the following regions of the parameter space:
\begin{eqnarray}
&&\theta_{M_1}=(-\pi,\pi),\;\theta_{\mu'}=(-\pi,\pi),\;\theta_{M_{BB'}}=(-\pi,\pi),\;\theta_{M_{B'}}=(-\pi,\pi).
\label{parameter space}
\end{eqnarray}
The allowed region of $\theta_{M_1}$, $\theta_{M_{BB'}}$ is displayed in Fig.~\ref{de-thetanew} (a), while the allowed region of $\theta_{M_1}$, $\theta_{M_{B'}}$ is displayed in Fig.~\ref{de-thetanew} (b). From the picture we can see that, when the possible cancellation happens between $\theta_{M_1}$ and new phases $\theta_{\mu'}$, $\theta_{M_{BB'}}$, $\theta_{M_{B'}}$ in the B-LSSM, the constraint from $d_e$ on $\theta_{M_1}$ can be relaxed completely. Comparing Fig.~\ref{de-thetanew} (a) with Fig.~\ref{de-thetanew} (b), it is obvious that the allowed region of $\theta_{M_1}$ versus $\theta_{M_{BB'}}$ as $\sin \theta_{M_{BB'}}$. But there is no obvious trend of the allowed region of $\theta_{M_1}$ with the changing of $\theta_{M_{B'}}$, which indicates that $\theta_{M_{BB'}}$ affects the numerical results more obviously than $\theta_{M_{B'}}$. Because $M_{BB'}$ is the mixing term between $\tilde\lambda_{B}$ and $\tilde\lambda_{B'}$, it contributes to $d_e$ through the channel of $\tilde\lambda_{B}$ and $\tilde\lambda_{B'}$, when $M_{B'}$ contributes to $d_e$ only through the channel of $\tilde\lambda_{B'}$.

\begin{figure}
\setlength{\unitlength}{1mm}
\centering
\includegraphics[width=3.1in]{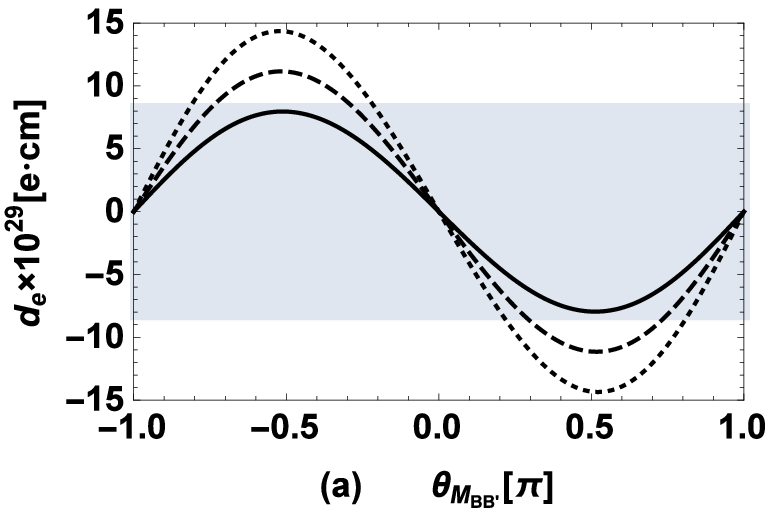}%
\vspace{0.5cm}
\includegraphics[width=3.1in]{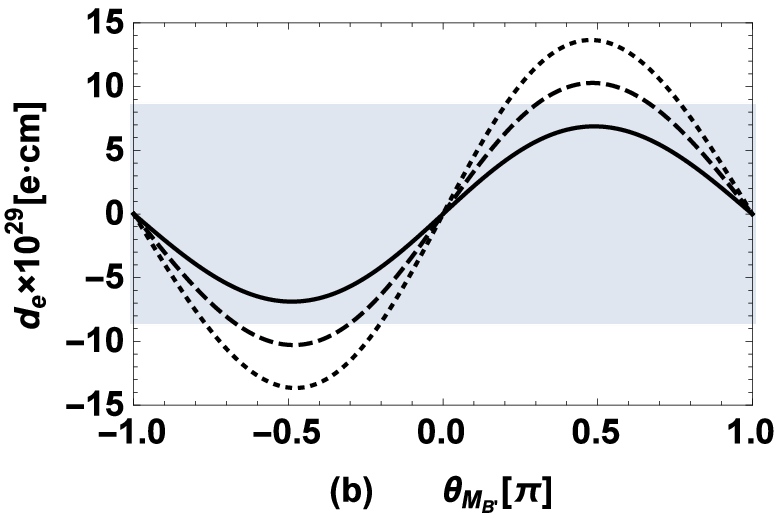}
\vspace{0cm}
\caption[]{$d_e$ versus $\theta_{M_{BB'}}$ (a) for $M_{BB'}=0.1\;{\rm TeV}$ (solid line), $0.14\;{\rm TeV}$ (dashed line), $0.18\;{\rm TeV}$ (dotted line), and $d_e$ versus $\theta_{M_{B'}}$ (b) for $M_{B'}=0.6\;{\rm TeV}$ (solid line), $0.9\;{\rm TeV}$ (dashed line), $1.2\;{\rm TeV}$ (dotted line), where the gray area denotes the present experimental upper bound on $d_e$.}
\label{de-newtheta}
\end{figure}
Assuming all contributions from other phases are cancelled each other completely, and the only contribution to $d_e$ comes from $\theta_{M_{BB'}}$. Then $d_e$ versus $\theta_{M_{BB'}}$ is plotted in Fig.~\ref{de-newtheta} (a), where the solid line, dashed line, dotted line denote $M_{BB'}=0.1\;{\rm TeV}$, $0.14\;{\rm TeV}$, $0.18\;{\rm TeV}$ respectively. Similarly, $d_e$ versus $\theta_{M_{B'}}$ is plotted in Fig.~\ref{de-newtheta} (b), where the solid line, dashed line, dotted line denote $M_{B'}=0.6\;{\rm TeV}$, $0.9\;{\rm TeV}$, $1.2\;{\rm TeV}$ respectively. The gray areas denote the present experimental upper bound on $d_e$. From the picture we can see that, $\theta_{M_{BB'}}$ or $\theta_{M_{B'}}$ affect the numerical results more obviously with the increasing of $M_{BB'}$ or $M_{B'}$, because the effects of $\theta_{M_{BB'}}$ or $\theta_{M_{B'}}$ are enlarged by large $M_{BB'}$ or $M_{B'}$. Comparing the effects of $M_{BB'}$ with $M_{B'}$, it is can be noted that $M_{BB'}$ affect the numerical results more obviously than $M_{B'}$, which coincides with the discussion of Fig.~\ref{de-thetanew} above.

\section{Summary\label{sec5}}

In this work, we focus on the CPV effects on EWB and electron EDM in the B-LSSM. Compared with the MSSM, new singlets mix with the MSSM doublets at the tree level, and a strong two-step PT can be achieved in this case. Moreover, new gaugino can make contributions to $Y_B$, and new coupling constant $g_{_{YB}}$ can affect the numerical results of $Y_B$ obviously. When the resonant enhancement $\mu\approx M_{1,B'}$ or $M_2$ take place, the minimum value of $\theta_\mu$ is about $0.03$ to give rise to EWB. In this case, the contributions to the electron EDM must be enhanced vastly, the cancellation of the contributions from different CPV phases is needed. In addition, the main contributions to electron EDM come from charginos, and $M_2$ which also appears in the chargino sector can make comparable contributions with the $\mu$ term. Hence, the contributions from $\theta_\mu$ to $d_e$ can be cancelled by $\theta_{M_2}$, and the worry about the contributions from the large value of $\theta_\mu$ to the electron EDM whether can be cancelled is relaxed. In addition, new CPV phases $\theta_{\mu'}$, $\theta_{M_{BB'}}$, $\theta_{M_{B'}}$ in the B-LSSM also can cancel the contributions from $\theta_{M_1}$ to $d_e$, and the allowed region of $\theta_{M_1}$ is relaxed completely when the cancellation between $\theta_{M_1}$ and $\theta_{\mu'}$, $\theta_{M_{BB'}}$, $\theta_{M_{B'}}$ takes place. Assuming that the only contributions to $d_e$ come from $M_{BB'}$ and $M_{B'}$, the numerical results show that, the experimental data of $d_e$ favor $M_{BB'}\lesssim0.1\;{\rm TeV}$, $M_{B'}\lesssim0.6\;{\rm TeV}$ when the regions of $\theta_{M_{BB'}}$ and $\theta_{M_{B'}}$ are relaxed completely.

\begin{acknowledgments}

The work has been supported by the National Natural Science Foundation of China (NNSFC) with Grants No. 11535002, and No. 11705045, Natural Science Foundation of Hebei province with Grants No. A2016201010, the youth top-notch talent support program of the Hebei Province, Hebei Key Lab of Optic-Eletronic Information and Materials, and the Midwest Universities Comprehensive Strength Promotion project.
\end{acknowledgments}


\begin{thebibliography}{99}
\bibitem{Cooke:2013cba}R. Cooke, M. Pettini, R. A. Jorgenson, M. T. Murphy and C. C. Steidel, Astrophys. J {\bf 781}, 31 (2014).
\bibitem{Ade:2015xua}P. A. R. Ade {\it et al.} [Planck Collaboration], Astron. Astrophys {\bf 594}, A13 (2016).
\bibitem{Kuzmin:1985mm}V. A. Kuzmin, V. A. Rubakov and M. E. Shaposhnikov, Phys. Lett. B {\bf 155}, 36 (1985).
\bibitem{Gavela:1981sk}M. B. Gavela, A. Le Yaouanc, L. Oliver, O. Pene, J. C. Raynal and T. N. Pham, Phys. Lett. B {\bf 109}, 215 (1982).
\bibitem{Bernreuther:1990jx}W. Bernreuther and M. Suzuki, Rev. Mod. Phys {\bf 63}, 313 (1991) Erratum: [Rev. Mod. Phys {\bf 64}, 633 (1992)].
\bibitem{Pospelov:2013sca}M. Pospelov and A. Ritz, Phys. Rev. D {\bf 89}, 056006 (2014).
\bibitem{de-1}J. Baron et al, (ACME Collaboration), Science {\bf343}, 269 (2014).
\bibitem{PDG}M. Tanabashi et al, (Particle Data Group), Phys. Rev. D {\bf98}, 030001 (2018).
\bibitem{de-3}V. Andreev et al, (ACME Collaboration), Nature {\bf562}, 355-360 (2018).
\bibitem{Dine:1990fj}M. Dine, P. Huet, R. L. Singleton, Jr and L. Susskind, Phys. Lett. B {\bf 257}, 351 (1991).
\bibitem{Cohen:1992zx}A. G. Cohen and A. E. Nelson, Phys. Lett. B {\bf 297}, 111 (1992).
\bibitem{Huet:1995sh}P. Huet and A. E. Nelson, Phys. Rev. D {\bf 53}, 4578 (1996).
\bibitem{Chang:2002ex}D. Chang, W. F. Chang and W. Y. Keung, Phys. Rev. D {\bf 66}, 116008 (2002).
\bibitem{Lee:2004we}C. Lee, V. Cirigliano and M. J. Ramsey-Musolf, Phys. Rev. D {\bf 71}, 075010 (2005).
\bibitem{Konstandin:2005cd}T. Konstandin, T. Prokopec, M. G. Schmidt and M. Seco, Nucl. Phys. B {\bf 738}, 1 (2006).
\bibitem{Chung:2008aya}D. J. H. Chung, B. Garbrecht, M. J. Ramsey-Musolf and S. Tulin, Phys. Rev. Lett. {\bf 102}, 061301 (2009).
\bibitem{Chung:2009qs}D. J. H. Chung, B. Garbrecht, M. J. Ramsey-Musolf and S. Tulin, JHEP {\bf 0912}, 067 (2009).
\bibitem{Chung:2009cb}D. J. H. Chung, B. Garbrecht, M. J. Ramsey-Musolf and S. Tulin, Phys. Rev. D {\bf 81}, 063506 (2010).
\bibitem{Cirigliano:2009yd}V ~Cirigliano, Y. Li, S. Profumo and M. J. Ramsey-Musolf, JHEP {\bf 1001}, 002 (2010).
\bibitem{Chiang:2009fs}C. W. Chiang and E. Senaha, JHEP {\bf 1006}, 030 (2010).
\bibitem{Morrissey:2012db}D. E. Morrissey and M. J. Ramsey-Musolf, New J. Phys. {\bf 14}, 125003 (2012).
\bibitem{Kozaczuk:2012xv}J. Kozaczuk, S. Profumo, M. J. Ramsey-Musolf and C. L. Wainwright, Phys. Rev. D {\bf 86}, 096001 (2012).

\bibitem{Pietroni:1992in}M. Pietroni, Nucl. Phys. B {\bf 402}, 27 (1993).
\bibitem{Davies:1996qn}A. T. Davies, C. D. Froggatt and R. G. Moorhouse, Phys. Lett. B {\bf 372}, 88 (1996).
\bibitem{Huber:2000mg}S. J. Huber and M. G. Schmidt, Nucl. Phys. B {\bf 606}, 183 (2001).
\bibitem{Kang:2004pp}J. Kang, P. Langacker, T. j. Li and T. Liu, Phys. Rev. Lett. {\bf 94}, 061801 (2005).
\bibitem{Huber:2006wf}S. J. Huber, T. Konstandin, T. Prokopec and M. G. Schmidt, Nucl. Phys. B {\bf 757}, 172 (2006).




\bibitem{Nath:1991dn}P. Nath, Phys. Rev. Lett {\bf 66}, 2565 (1991).
\bibitem{Kizukuri:1992nj}Y. Kizukuri and N. Oshimo, Phys. Rev. D {\bf 46}, 3025 (1992).
\bibitem{Falk:1996ni}T. Falk and K. A. Olive, Phys. Lett. B {\bf 375}, 196 (1996).
\bibitem{Falk:1998pu}T. Falk and K. A. Olive, Phys. Lett. B {\bf 439}, 71 (1998).
\bibitem{Brhlik:1998zn}M. Brhlik, G. J. Good and G. L. Kane, Phys. Rev. D {\bf 59}, 115004 (1999).
\bibitem{Bartl:1999bc}A. Bartl, T. Gajdosik, W. Porod, P. Stockinger and H. Stremnitzer, Phys. Rev. D {\bf 60}, 073003 (1999).
\bibitem{Abel:2001vy}S. Abel, S. Khalil and O. Lebedev, Nucl. Phys. B {\bf 606}, 151 (2001).
\bibitem{Barger:2001nu}V. D. Barger, T. Falk, T. Han, J. Jiang, T. Li and T. Plehn, Phys. Rev. D {\bf 64}, 056007 (2001) [hep-ph/0101106].
\bibitem{Olive:2005ru}K. A. Olive, M. Pospelov, A. Ritz and Y. Santoso, Phys. Rev. D {\bf 72}, 075001 (2005) [hep-ph/0506106].
\bibitem{Cirigliano:2006dg}V. Cirigliano, S. Profumo and M. J. Ramsey-Musolf, JHEP {\bf 0607}, 002 (2006) [hep-ph/0603246].
\bibitem{YaserAyazi:2006zw}S. Yaser Ayazi and Y. Farzan, Phys. Rev. D {\bf 74}, 055008 (2006).
\bibitem{Engel:2013lsa}J. Engel, M. J. Ramsey-Musolf and U. van Kolck, Prog. Part. Nucl. Phys. {\bf 71}, 21 (2013).
\bibitem{Chupp:2017rkp}T. Chupp, P. Fierlinger, M. Ramsey-Musolf and J. Singh,Rev. Mod. Phys. {\bf 91}, 015001 (2019).
\bibitem{Barger:2008wn}V. Barger, P. Fileviez Perez and S. Spinner, Phys. Rev. Lett {\bf 102}, 181802 (2009).
\bibitem{FileviezPerez:2008sx}P. Fileviez Perez and S. Spinner, Phys. Lett. B {\bf 673}, 251 (2009).
\bibitem{5}M. Ambroso and B. A. Ovrut, Int. J. Mod. Phys. A {\bf 26}, 1569 (2011).
\bibitem{6}P. F. Perez and S. Spinner, Phys. Rev. D {\bf83}, 035004 (2011).
\bibitem{16}S. Khalil and H. Okada, Phys. Rev. D {\bf79}, 083510 (2009).
\bibitem{1616}L. Basso, B. O¡¯Leary, W. Porod and F. Staub, JHEP {\bf1209}, 054 (2012).
\bibitem{DelleRose:2017ukx}L. Delle Rose, S. Khalil, S. J. D. King, C. Marzo, S. Moretti and C. S. Un, Phys. Rev. D {\bf 96}, 055004 (2017).
\bibitem{DelleRose:2017uas}L. Delle Rose, S. Khalil, S. J. D. King, S. Kulkarni, C. Marzo, S. Moretti and C. S. Un, JHEP {\bf 1807}, 100 (2018).
\bibitem{search}W. Abdallah, A. Hammad, S. Khalil and S. Moretti, Phys. Rev. D {\bf 95}, 055019 (2017).
\bibitem{77}A. Elsayed, S. Khalil and S. Moretti, Phys. Lett. B {\bf715}, 208 (2012).
\bibitem{88}G. Brooijmans et al. [arXiv:1203.1488 [hep-ph]].
\bibitem{9}L. Basso and F. Staub, Phys. Rev. D {\bf87}, 015011 (2013).
\bibitem{99}L. Basso et al., Comput. Phys. Commun. {\bf184}, 698 (2013).
\bibitem{10}A. Elsayed, S. Khalil, S. Moretti and A. Moursy, Phys. Rev. D {\bf87}, 053010 (2013).
\bibitem{11}S. Khalil and S. Moretti, Rept. Prog. Phys {\bf80}, 036201 (2017).
\bibitem{164}F. Staub, arXiv:0806.0538. F. Staub, Comput. Phys. Commun. {\bf181} 1077-1086 (2010). F. Staub, Comput. Phys. Commun. {\bf182} 808-833 (2011). F. Staub, Comput. Phys. Commun. {\bf184} 1792-1809 (2013). F. Staub, Comput. Phys. Commun. {\bf185} 1773-1790 (2014).
\bibitem{Yang:2018utw}J. L. Yang, T. F. Feng, H. B. Zhang, G. Z. Ning and X. Y. Yang, Eur. Phys. J. C {\bf 78}, 438 (2018).
\bibitem{JLYang:2018}J. L. Yang, T. F. Feng, S. M. Zhao, R. F. Zhu, X. Y. Yang and H. B. Zhang, Eur. Phys. J. C {\bf 78}, 714 (2018).
\bibitem{Yang:2018guw}J. L. Yang, T. F. Feng, Y. L. Yan, W. Li, S. M. Zhao and H. B. Zhang, Phys. Rev. D {\bf 99}, 015002 (2019).
\bibitem{Delepine:1996vn}D. Delepine, J. M. Gerard, R. Gonzalez Felipe and J. Weyers, Phys. Lett. B {\bf 386}, 183 (1996).
\bibitem{Carena:1996wj}M. Carena, M. Quiros and C. E. M. Wagner, Phys. Lett. B {\bf 380}, 81 (1996).
\bibitem{Espinosa:1996qw}
  J. R. Espinosa, Nucl. Phys. B {\bf 475}, 273 (1996).
\bibitem{Carena:1997ki}M. Carena, M. Quiros and C. E. M. Wagner, Nucl. Phys. B {\bf 524}, 3 (1998).
\bibitem{Huber:1998ck}S. J. Huber and M. G. Schmidt, Eur. Phys. J. C {\bf 10}, 473 (1999).
\bibitem{Carena:2002ss}M. Carena, M. Quiros, M. Seco and C. E. M. Wagner, Nucl. Phys. B {\bf 650}, 24 (2003).
\bibitem{Profumo:2007wc}S. Profumo, M. J. Ramsey-Musolf and G. Shaughnessy, JHEP {\bf 0708}, 010 (2007).
\bibitem{Carena:2008vj}M. Carena, G. Nardini, M. Quiros and C. E. M. Wagner, Nucl. Phys. B {\bf 812}, 243 (2009).
\bibitem{Curtin:2012aa}D. Curtin, P. Jaiswal and P. Meade, JHEP {\bf 1208}, 005 (2012).
\bibitem{Krizka:2012ah}K. Krizka, A. Kumar and D. E. Morrissey, Phys. Rev. D {\bf 87}, 095016 (2013).
\bibitem{Patel:2011th}H. H. Patel and M. J. Ramsey-Musolf, JHEP {\bf 1107}, 029 (2011).

\bibitem{Zhang:2013hva}H. B. Zhang, T. F. Feng, S. M. Zhao and T. J. Gao, Nucl. Phys. B {\bf 873}, 300 (2013) [Erratum: Nucl. Phys. B {\bf 879}, 235 (2014)].
\bibitem{Zhang:2013jva}H. B. Zhang, T. F. Feng, G. H. Luo, Z. F. Ge and S. M. Zhao, JHEP {\bf 1307}, 069 (2013) [Erratum: JHEP {\bf 1310}, 173 (2013)].
\bibitem{Yang:2009zzh}X. Y. Yang and T. F. Feng, Phys. Lett. B {\bf 675}, 43 (2009).
\bibitem{Yang:2019aao}J. L. Yang, T. F. Feng, S. K. Cui, C. X. Liu, W. Li and H. B. Zhang, arXiv:1910.05868 [hep-ph].
\bibitem{HiggsC1}M. Carena, J. R. Espinosaos and C. E. M. Wagner, M. Quir, Phys. Lett. B {\bf355}, 209 (1995).
\bibitem{HiggsC2}M. Carena, M. Quiros and C. E. M. Wagner, Nucl. Phys. B, {\bf461}, 407 (1996).
\bibitem{HiggsC3}M. Carena, S. Gori, N.R. Shah and C. E. M. Wagner, JHEP, {\bf03}, 014 (2012).
\bibitem{newZ}ATLAS Collab., ATLAS-CONF-2016-045.
\bibitem{20}G. Cacciapaglia, C. Csaki, G. Marandella, and A. Strumia, Phys.Rev. D {\bf74}, 033011 (2006) [hep-ph/0604111] .
\bibitem{21}M. Carena, A. Daleo, B. A. Dobrescu and T. M. P. Tait, Phys. Rev. D {\bf70}, 093009 (2004).
\bibitem{AlvarezGaume:1983gj}L. Alvarez-Gaume, J. Polchinski and M. B. Wise, Nucl. Phys. B {\bf 221}, 495 (1983).
\bibitem{Strumia:1996pr}A. Strumia, Nucl. Phys. B {\bf 482}, 24 (1996).
\bibitem{Profumo:2006yu}S. Profumo, M. J. Ramsey-Musolf and S. Tulin, Phys. Rev. D {\bf 75}, 075017 (2007).

\bibitem{Moreno:1998bq}J. M. Moreno, M. Quiros and M. Seco, Nucl. Phys. B {\bf 526}, 489 (1998).
\bibitem{John:2000zq}P. John and M. G. Schmidt, Nucl. Phys. B {\bf 598}, 291 (2001) Erratum: [Nucl. Phys. B {\bf 648}, 449 (2003)].
\bibitem{Akula:2017yfr}S. Akula, C. Bal¨¢zs, L. Dunn and G. White, JHEP {\bf 1711}, 051 (2017).
\bibitem{Carena:1997gx}M. Carena, M. Quiros, A. Riotto, I. Vilja and C. E. M. Wagner, Nucl. Phys. B {\bf 503}, 387 (1997).
\bibitem{Carena:2000id}M. Carena, J. M. Moreno, M. Quiros, M. Seco and C. E. M. Wagner, Nucl. Phys. B {\bf 599}, 158 (2001).






\end{thebibliography}
\end{document}